\documentclass[aps,prd,twocolumn,groupedaddress,showpacs]{revtex4}
\usepackage{graphicx}% Include figure files
\usepackage{dcolumn}% Align table columns on decimal point
\usepackage{bm}% bold math
\usepackage{amssymb,amsmath,latexsym,amsfonts}
\begin{document}

\title{Effective Quantum Mechanics of a falling particle}

\author{Guillermo Chac\'{o}n-Acosta}
\email{gchacon@correo.cua.uam.mx} \affiliation{Departamento de
Matem\'aticas Aplicadas y Sistemas, \\ 
Universidad Aut\'onoma Metropolitana-Cuajimalpa,\\
Vasco de Quiroga 4871,  Ciudad de M\'exico, 05348, Mexico}

\author{H\'ector Hugo Hern\'andez-Hern\'andez}
\email{hhernandez@uach.mx} \affiliation{ Facultad de Ingenier\'ia\\Universidad Aut\'onoma de Chihuahua\\
Nuevo Campus Universitario, Chihuahua 31125, M\'exico.}

\author{ Mercedes Vel\'azquez}
\email{mquesada@fcfm.buap.mx}
\affiliation{Facultad de Ciencias F\'isico Matem\'aticas, \\ 
Benem\'erita Universidad Aut\'onoma de Puebla,\\
Puebla 72570,  Puebla, 05348, Mexico}

\begin{abstract}
We analyze the problem of one dimensional quantum particle falling in a constant gravitational field, also known as the {\it bouncing ball}, employing a semiclassical approach known as momentous effective quantum mechanics. In this formalism the quantum evolution is described through a dynamical system of infinite dimension for the position, the momentum and all dispersions. Usually the system is truncated to have a finite dimensional one, however, in this case equations of motion decouple and the system can be solved. For a specific set of initial conditions we find that the time dependent dispersion in position is always around the classical trajectory.
\end{abstract}

%\date{\today}
%\pacs{04.60Bc, 04.60.Kz, 04.60.Pp, 03.75.Nt
\maketitle

\section{Introduction}
The description of classical and quantum phenomena has in general, profound differences in its mathematical formulation and its physical interpretation. On the one hand, in 
classical mechanics one studies the state of physical systems by analyzing the movement of its constituents under the influence of external forces, or its energy (Hamiltonian) description, and the time evolution of such systems is based on a set of ordinary differential equations of motion. In such a way, once the physical variables describing the system are obtained as functions of time, for a set of initial conditions, one knows a priori the behavior of the system at all times, hence classical mechanical systems are completely deterministic.
On the other hand, quantum systems have a probabilistic description, for instance, it is not possible to know simultaneously the position and momentum of a given particle at a given time, according to Heisenberg's uncertainty principle, therefore the concept of {\it trajectory} is lacking. All the information of the system is codified in the wave function $\psi(\mathbf{x},t)$, whose squared norm gives the probability of the system of being in a certain state. The evolution of the wave function is determined by solving the well-known Schr\"odinger equation.

Given the probabilistic nature of the quantum mechanical description of physical systems, its interpretation is not always straightforward, being also complicated because the solution of the Schr\"odinger equation involves boundary conditions, even for the simplest systems. Hence, the use of approximation methods is not only desirable but necessary in order to study such systems and to better understand their behavior. There exist several such methods, most notable the WKB and perturbation methods \cite{dynamics_book}. It is possible however, under certain assumptions, to give an equivalent description to the quantum evolution by means of a {\it quasi classical} description, on the grounds that quantum mechanics should reduce to a classical theory in limit when $\hbar \rightarrow 0$. Indeed, the so-called Ehrenfest theorem says that, for a sufficiently narrow wave package, the expected value of the position in that  state follow on average a classical trajectory. However, an infinite number of corrections around such classical trajectory are needed to reconstruct the corresponding quantum state \cite{Balle}. Hence, in order to describe the behavior and evolution of a quantum system by means of a set of quasi classical differential equations, evolution equations for quantum corrections must be included. 

Semiclassical effective quantum mechanics is an approach that has been developed to systematically analyze quantum effects through modifications of the classical equations, leading to observable effects as deviations from the classical behavior \cite{Bojo1}. This approach allow us to approximate the evolution of the quantum state of the physical system through a hierarchy of coupled classical dynamical equations for configuration variables and for quantum dispersions or uncertainties, that in turn and under certain conditions, can be considered corrections that slightly modify the classical behavior. The method not only serves as an approximation to the quantum behavior but it is equivalent to the latter when all the infinitely many quantum corrections are taken into account. It also proves useful when the concept of {\emph quantum trajectory} is needed because semiclassical trajectories can be generated in this setting.

In this work we apply the momentous effective formalism to a quantum particle in a the presence of a gravitational potential. For this problem it is shown that the effective system reduces to a finite number of quantum corrections, thus making the system exactly soluble. In section \ref{quantum} we review the classical and quantum dynamics of the quantum ball. In section \ref{semiclassical} we describe the method of effective equations for the moments, i.e. classical and quantum variables, analyzing the quantum gravitational effect on the bouncing ball and discuss its deviation from the classical behavior due to quantum corrections.

\section{Quantum particle in a gravitational potential and scaling}\label{quantum}

\subsection{Classical bounce}
The dynamics of a particle under the influence of a gravitational potential $V(x) = mgx$ is governed by the Hamiltonian
\begin{equation}\label{mgh}
H(x,p) = \frac{p^2}{2m} + mgx.
\end{equation}
For general initial conditions $(x(t), \dot{x}(t) )|_{t=0} = (x_0 , v_0)$, the solution of the equations of motion for the position $x$ as a function of time is 
\begin{equation}\label{Sol}
x(t) = x_0+v_0t-\frac{g}{2}t^2,
\end{equation}
where $g$ is the acceleration due to gravity. As aforementioned Eq. (\ref{Sol}) determines the position of the particle for any time $t> 0$.
Without any lose of generality we can choose $v_0=0$, a different value of the initial velocity simply shifts the value of $x_{0}$.

As in \cite{Julio}, we place a mirror at the bottom of the potential so the system is conservative and we have infinite parabolas, one for each bounce. It is possible to write the general solution in the following way in terms of the period 
\begin{eqnarray}\label{class_sol} 
 x(t)&=& \left( x_{0}- \frac{g}{2}t^2 \right) \Theta(t) + \nonumber \\ 
&+& 2gT \sum_{n=1}^{\infty} \left( t- (2n-1)T \right) \Theta\left( t- (2n-1)T \right),
\end{eqnarray}
where $T^{2}=2 x_0 / g$ is the drop time and $\Theta(t)$ is the Heaviside step function. This solution can be expanded in Fourier series in order to be compared with the quantum case \cite{Julio}, giving
\begin{equation}
    x(t) \approx \frac{2}{3} x_{0} + \frac{4 x_{0}}{\pi^{2}} \sum_{n=1}^{\infty} \frac{(-1)^{n} }{n^{2}}  \cos{\left( \frac{2 \pi n}{T}t \right)} .
\end{equation}
%%%%%%%%%%%

\subsection{Quantum bounce}

Now we study the quantum behavior of a massive quantum particle in the presence of this constant gravitational field. The motion of a falling particle above a mirror is known as the quantum bouncer \cite{Julio,JGBWKB}, the reflective potential potential can be simulated by solving the Schr\"odinger equation in the region $x>0$. For neutrons the characteristic length scale of the system is of the order of peV, an energy scale that can be used in measurements of fundamental constants and gravitational effects beyond Newtonian theory \cite{new}.

For this gravitational potential $V ( x) = mg x$, the quantum Hamiltonian operator is
\begin{equation} \label{Schroedinger}
\hat{H} = \frac{\hat{p}^{2}}{2m} + mg \hat{x} = - \frac{\hbar^{2}}{2m}\frac{\partial^{2}}{\partial x^{2}} + mg x,
\end{equation}
whose eigenvalue equation is
\begin{equation}\label{propia}
E_n\psi_n= - \frac{\hbar^{2}}{2m}\frac{\partial^{2}\psi_n}{\partial x^{2}} + mg x\psi_n.
\end{equation}
 $\psi_n(x)$ are the wave eigenfunctions and $E_n$ the energy spectrum.
Equation (\ref{propia}) can be rescaled to be dimensionless by introducing a characteristic length and energy scales 
\begin{eqnarray}
l_g & = & \left(\frac{\hbar^{2}}{2g m^{2}} \right)^{1/3}, \label{lg} \\
E_g & = & \left( \frac{\hbar^2 g^2 m}{2} \right)^{1/3} = mg l_g\,.  \label{Eg}
\end{eqnarray}
Thus, with the new variables $x^{\star} = \frac{x}{l_{g}}$ and $E^{ \star}= \frac{E}{mg l_{g}}$, Eq. (\ref{propia}) turns into an Airy equation of the form $\psi''(x^{\star})=(x^{\star}-E^{\star})\psi(x^{\star})$, whose solution are the Airy functions shifted from the roots $x_n$ as follows
\begin{equation} \label{psi_{Airy}}
\psi_n (x^{\star}) = N_n \text{Ai}(x^{\star} - x_n),
\end{equation}
where the $N_n$ are normalization constants.

The Airy function is defined by
\begin{equation}
\text{Ai} (x) = \frac{1}{\pi}\int\limits_{0}^{\infty}\cos\left(\frac{t^{3}}{3} + xt\right) dt.
\end{equation}
The energy spectrum is written in terms of the zeros of the Airy function, $x_n > 0$, such that 
 $E_n = mgl_gx_n $, for $n = 1,2,\dots$
An analytical approximation of $x_n$ for large $n$ is given in \cite{siam}, yielding for the energies
\begin{equation}\label{espectro}
E_n \cong mg l_g\left[\frac{3\pi}{2}\left(n - \frac{1}{4}\right)\right]^{2/3}.
\end{equation}
The first eigenvalue for $n = 1$ is equal to $ E/(mg l_g) = 2\text{.}33811$, while the approximation (\ref{espectro}) gives the value of $E/(mg l_g) \simeq 2\text{.}32025$, which only differs from the former by $0\text{.}76372\%$. 
For neutrons with mass $940 \, \text{MeV}/c^2$ the corresponding gravitational length is $l_g=5.87\mu \text{m}$, whose corresponding energy scale is $E_g=0\text{.}602 \, \text{peV}$, \cite{new}. Recently, experiments showing that the transitions between the quantum states of neutrons can be triggered by the vibrations of a mirror have been proposed. In such cases the quantum bouncer is used for measurements of sensitivity in gravity tests \cite{nat,seveso}.

An approximation of the evolution of the expectation value for position in (\ref{Schroedinger}) with gaussian initial wave packet $\Psi(x,0)= \left( {2}/{\pi \sigma^{2}} \right)^{1/4} e^{-(x-x_{0})^2 /\sigma^2}$, where $\sigma$ is its width and is height is $x_0$,  can be given for high quantum energies, corresponding to the semiclassical limit, and is as follows
\begin{eqnarray} \label{expectation_x}
 \langle x(t) \rangle &\approx& \sum_{n=-\infty}^{\infty} e^{- \frac{\pi^{2} n^{2}}{2 \sigma^{2} } x_{0} } A_{n} e^{ \frac{-2 i n \pi }{T} t} \nonumber \\
 &=&  \frac{2}{3} x_{0} + \frac{4 x_{0}}{\pi^{2}} \sum_{n=1}^{\infty} \frac{(-1)^{n} }{n^{2}} e^{- \frac{\pi^{2} n^{2}}{2 \sigma^{2} } x_{0} } \cos{\left( \frac{2 \pi n}{T}t \right)}.
\end{eqnarray}
Note that for large height $x_0$ the first factor in the sum (\ref{expectation_x}) tends to unity and the Fourier series of (\ref{class_sol}) is retrieved. 

As shown in \cite{Julio}, for certain initial states the expectation values of the position bounce, collapse to the classical value for some time, and then revive and start to bounce again. The revivals can  be interpreted as an interference between different sections of the wave packet that have already bounced. This system has also been approached with the WKB approach with which is the possible effect of tunneling resonances \cite{JGBWKB}. Further, it has been found that a falling wave packet can have a diffractive structure depending on its initial dispersion, if it is wider than the characteristic length $l_g$, it falls as a free packet, but if it is smaller, it has a diffraction pattern \cite{JPAMG}. Certainly, the appearance of such collapses and revivals was visualized on light propagation in an optical waveguide \cite{owg}. This model has been further extended to matter-wave soliton that bounces more like a particle compared to the wave packet \cite{Benseghir}. 

%%%%%%%%%%%%%%%%%%%%%%%%%
%%%%%%%%%%%%%%%%%%%%%%%%%%

\section{Momentous effective formulation of the quantum falling particle} \label{semiclassical}

As it is well known it is impossible to know simultaneously the position and momentum of a quantum particle or system due to the Heisenberg uncertainty principle, that is, there exist no quantum trajectories analogous to the classical evolution.  However, as we mentioned above, there is a particular semiclassical formalism in which a quantum mechanical system can be studied as a classical Hamiltonian system but in a infinite dimensional phase space, where the additional degrees of freedom are directly related to the expectation values of all the infinite many quantum dispersions \cite{Bojo1}. This formalism was first applied to soluble models of loop quantum cosmology  to show the existence of a quantum bounce at the beginning of the universe \cite{BojoDCS,QC,Bojo,DBriz}. However, it has also been studied in other systems such as anharmonic oscillators \cite{Briz}, and also two dimensional systems as the quantum Kepler problem \cite{Hdz}. The name of {\it momentous} effective quantum dynamics comes from the similarity of the quantum variables, that are the expected values of the dispersions, with the statistical moments of the probability distributions, although the former also contain additional quantum effects from the non commutativity of the quantum operators \cite{Briz1}. One of the very interesting features of such formulation is that it yields a classical dynamical evolution that is,  one can obtain trajectories for the quantum evolution provided with suitable initial conditions, laying an ideal ground for comparison between different analysis.

For a system with one degree of freedom the expectation value of quantum dispersions or momenta are defined as follows
\begin{equation}
G^{a, b} = \left\langle(\hat{p}-p)^{a}(\hat{x}- x)^{b}\right\rangle,
\end{equation}
where, the completely symmetric or Weyl order of the operators is used.

The evolution of the system is generated by an effective Hamiltonian $H_{Q}$, which is corrected form the classical one by quantum effects induced by the expectation values of dispersions, and is obtained through a Taylor expansion around the expectation values of the canonical variables
\begin{eqnarray}
H_{Q} \left(x, p, G^{a, b}\right) &=& \langle\hat{H}(\hat{x},\hat{p})\rangle\\
&= &H(x,p) + \sum\limits_{a+b\geq 2}^{\infty}\frac{1}{a! b!}\frac{\partial^{a+b}H(x,p)}{\partial p^{a}\partial x^{b}} G^{a,b}.\label{Hq0}\nonumber
\end{eqnarray}
where $p = \langle\hat{p}\rangle$ and $x = \langle\hat{x}\rangle$ are the expectation values of the momentum and position operators, and  $a, b=0,1,2,\ldots$, such that $\ a+b\geq 2$.

For the gravitational potential we get the following
\begin{equation}\label{Hq}
H_{Q} \left(x, p,G^{2,0}\right) = \frac{p^{2}}{2m} + mg x + \frac{1}{2m}G^{2,0}.
\end{equation}
An interesting feature of this system is that the Hamiltonian is finite in quantum corrections, just $G^{0,2}$ appears in the previous equation, so we can obtain its exact description.

The momenta $G^{a,b}$, being the expectation values of generic products of momentum and position operators must satisfy generalized uncertainty relations, given that the expected value of the operators squared satisfies Schwarz inequality \cite{BojoDCS}
\begin{equation}\label{GUP}
G^{0,2}G^{2,0} - (G^{1,1})^{2} \geq \frac{1}{4}\langle -i[\hat{x},\hat{p}]\rangle^2 =   \left(\frac{\hbar}{2}\right)^{2}.
\end{equation}

Equations of motion, for momenta and classical phase variables, are obtained with the Poisson brackets with the effective Hamiltonian (\ref{Hq0}),  \cite{Bojo1,BojoDCS}
\begin{eqnarray}
\frac{dx}{dt} & = & \{x, H_{Q}\}  =  \frac{\partial H}{\partial p} \nonumber \\
 &+& \sum\limits_{a+b\geq 2}^{\infty}\frac{1}{a! b!}\frac{\partial^{a+b+1}H(x,p)}{\partial p^{a+1}\partial x^{b}} G^{a,b}, \label{xq}
\end{eqnarray}
\begin{eqnarray}
\frac{dp}{dt} & = & \{p, H_{Q}\}  = - \frac{\partial H}{\partial x} \nonumber \\
 &-& \sum\limits_{a+b\geq 2}^{\infty}\frac{1}{a! b!}\frac{\partial^{a+b+1}H(x,p)}{\partial p^{a}\partial x^{b+1}} G^{a,b}, \label{pq}
\end{eqnarray}
\begin{eqnarray}
\frac{dG^{a,b}}{dt} & = & \{G^{a,b}, H_{Q}\}  =  \frac{b}{m}G^{a+1, b-1} \nonumber \\
 &+& a \sum\limits_{n=2}^{\infty}\frac{V^{(n)}(x)}{(n-1)!}\left[G^{0,n-1}G^{a-1,b} - G^{a-1, b+n-1}\right], \label{Gq}
\end{eqnarray}

In the case of Hamiltonian (\ref{Hq}) the equations of motion for expectation values $x,p,$ reduces to the classical Hamiltonian equations because the sums in (\ref{xq}) and (\ref{pq}) will give no contribution, indeed this occurs if the effective Hamiltonian is at most quadratic in canonical variables. Therefore the motion equations are the following
\begin{eqnarray}\label{EcsHamilton}
\frac{dG^{0,2}}{dt} &=& \frac{2}{m}G^{1,1},\quad \frac{dG^{1,1}}{dt} = \frac{1}{m}G^{2,0}, \quad \frac{dG^{2,0}}{dt} = 0, \\
\dot{x} &=& \frac{p}{m}, \quad \dot{p}=-mg.
\end{eqnarray}
As can be seen classical and quantum variables decouple. It is straightforward to solve them, and we obtain
\[G^{2,0}(t) = c_0 =const., \quad G^{1,1}(t) = \frac{c_0}{m}t + c_1, \]
\begin{equation}\label{Gs}
G^{0,2}(t) =  \frac{c_0}{m^2}t^2 + \frac{2 c_1}{m}t +c_2,  
\end{equation}
where the classical solution for the canonical variables in the same as (\ref{Sol}) and $c_0$, $c_1$ y $c_2$ are constants fixed by initial conditions.

We can use the solution (\ref{Gs}) in (\ref{GUP}) to get
\begin{equation}\label{GUP2}
c_0\,c_2- c_1^{2} \geq \left(\frac{\hbar}{2}\right)^{2}.
\end{equation}
Assuming that the dispersions are not initially correlated $G^{1,1}(0)=0$, implies that  $c_1=0$. The inequality depends only on  $c_0$ y $c_2$ which are the initial conditions for second order moments. Even more, if we saturate the uncertainty condition (\ref{GUP2}), we get a relation between constants fixed by initial conditions, that is
\begin{equation}\label{GUP3}
 c_0=  \frac{\hbar^2}{4c_2}.
\end{equation}
This set of initial conditions is certainly one of the simplest choices one can make, however, it is possible to make an analysis for coherent dynamical states for this model, in a similar way as the Gaussian-Klauder coherent states studied in \cite{edoGK}.

We choose to obtain the dispersion in positions since it is the one appearing in the corrected Hamiltonian (\ref{Hq}). If we want a non-trivial dynamical evolution for this momentum its initial condition must be non zero.
As we recall, there exists a characteristic length for the gravitational field $l_g$, given in Eq. (\ref{lg}), we can propose an initial condition for the moment of that order, i.e.

$$
G^{0,2}(0)= c_2=\alpha l_g^2,
$$
where $\alpha$ is a dimensionless constant. Using (\ref{GUP3}) we obtain  $c_0=(\hbar g m^2)^{2/3}/4\alpha$.
From this we see that the solutions in (\ref{Gs}) can be written as
\[G^{2,0}(t) = \frac{1}{4\alpha}(\hbar g m^2)^{2/3}, \quad G^{1,1}(t) = \frac{1}{4\alpha}(\hbar^2 g^2 m)^{1/3}\,t , \]
\begin{equation}\label{Gs1}
G^{0,2}(t) =    \frac{1}{4\alpha}  \left(\frac{\hbar g}{m}\right)^{2/3} t^2 +\alpha l_g^2,
\end{equation}
from where we see that the uncertainty in the momentum is constant, in position is quadratic and the correlation between both is linear in time. Substituting this last expression in  the Hamiltonian (\ref{Hq}), we can  rewrite it as follows
\begin{equation}\label{Hq2}
H_{Q} \left(x, p\right) = \frac{p^{2}}{2m} + mgx + \frac{1}{\alpha} \left( \frac{\hbar^2 g^2 m}{2} \right)^{1/3},
\end{equation}
where the correction term in the Hamiltonian is precisely proportional to the characteristic energy of the system $E_g$, given in (\ref{Eg}). This suggest that one option is to choose  $\alpha_1=1$ such that initial dispersion in the position might be the gravitational length squared. Next we will see a different choice, so let us keep $\alpha$ general, so that
\begin{equation}\label{Hq3}
H_{Q} \left(x, p\right) = \frac{p^{2}}{2m} + mg\left(x+\frac{l_g}{\alpha}\right).
\end{equation}
Let us realize that this Hamiltonian has a minimum non-zero energy associated with the characteristic length of the system. Indeed, it is possible to use the lowest energy eigenvalue to fix the lower bound of this Hamiltonian, so set $\alpha_2 = 0\text{.}4277$ as a different value for the constant of the position variance.

Thus, the expectation value for the position approximately follows the classical trajectory (\ref{Gs})  being bounded by the time-dependent uncertainty (\ref{Gs1}), so we can express the corresponding solution with effective quantum correction as $x(t)\pm \sqrt{G^{0,2}(t)}$, constrained to the same initial condition $v_0=0$, i.e.
\begin{equation}\label{xG}
 x_0-\frac{g}{2}t^2 \pm \sqrt{\alpha l_g^2 + \frac{g l_g}{2^{5/3}\alpha} t^2 }.
\end{equation}
\begin{figure}[t!]
 \includegraphics[width=0.5\textwidth]{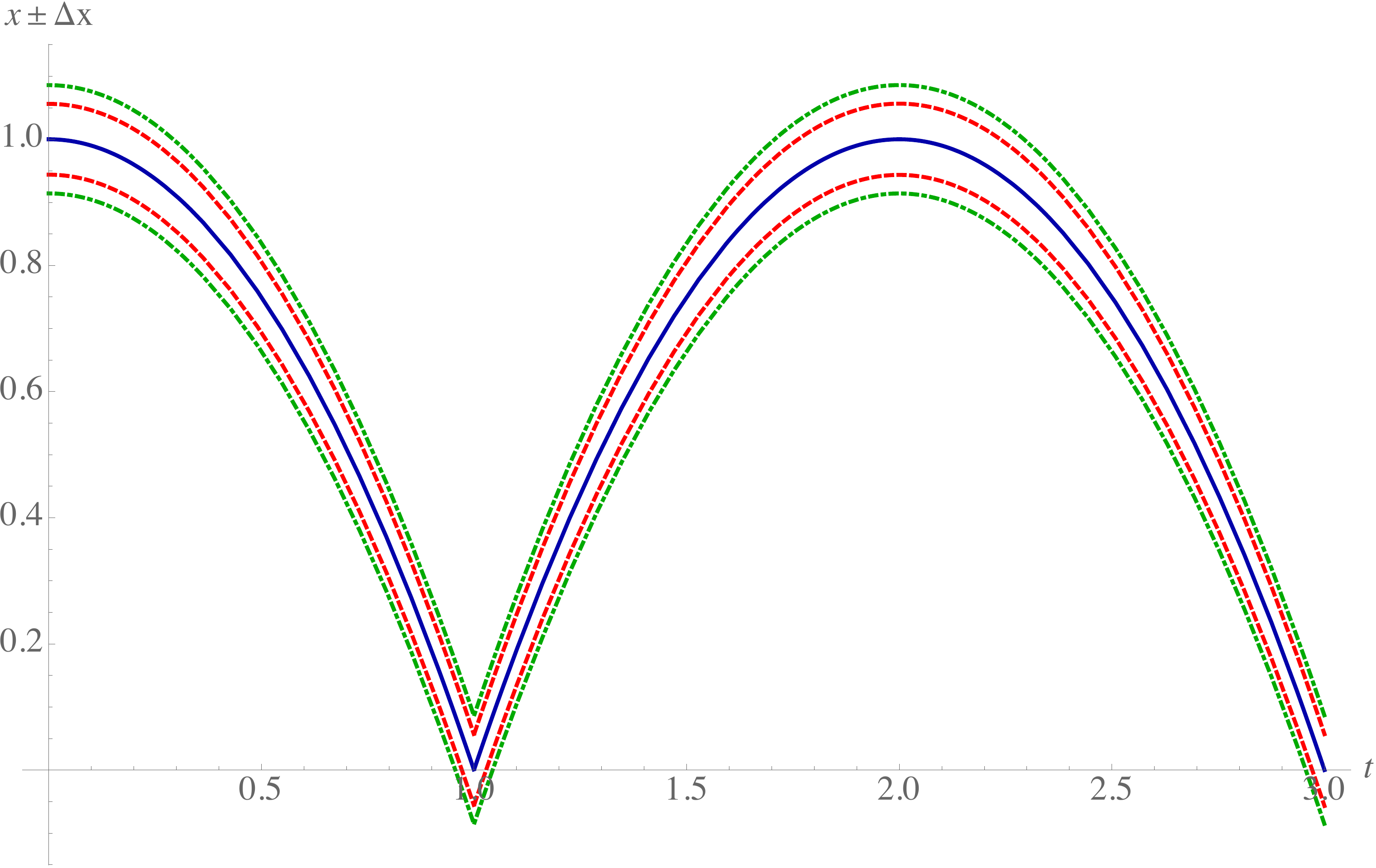}
 \caption{Semiclassical bounce of the quantum ball $x\pm \Delta x$. Solid (blue) line is the classical solution, dashed (red) line is for $\alpha_1=1$, and dot-dashed line (green) is the case $\alpha_2=0\text{.}4277$. } \label{1bounce}
\end{figure}
% Even though the error grows with time, the relative uncertainty $\sqrt{G^{0,2}}/x$ tends to zero for large $t$.

As in (\ref{class_sol}) it is possible to write the solution in terms of the period obtaining a series of bounces bounded by the uncertainty $\Delta x$. The behavior of Eq. (\ref{xG}) is shown in Fig. \ref{1bounce} the proposed values of the dispersion dimensionless constant $\alpha$. We realize that at the maximum of the trajectory, the absolute value of the uncertainty is a maximum. However, the behavior of the dispersions is the same in each period. The case of the momentum is very simple since the uncertainty is always constant at each point, so the graph will be a straight line with a constant fringe. Finally, the covariance is linear in $t$, which means that as we evolve in time the dispersions of classical variables will be more correlated up to a maximum in each period, due to quantum effects. Indeed the strength of the correlation is proportional to the characteristic energy $E_g$.

%%%%%%%%%%%%%%%%%%%%%%%%%%%

\section{Discussion}

We have employed the momentous effective semiclassical description of quantum mechanics to study the behavior of a particle under the influence of a gravitational potential. We showed how the usual Schr\"odinger evolution in quantum mechanics can be analyzed with an equivalent semiclassical dynamical system for which a Hamiltonian is obtained. The dynamical phase space variables of this system, ($x$, $p$) together with all quantum dispersions  $G^{a,b}$ encode all the information of the quantum system, making the study of quantum systems more tractable for a particular truncation.

The quantum behavior of a massive  particle bouncing in the presence of a gravitational field, which is usually called a quantum bouncer, is used to model some experiments to measure sensibility in gravity tests, making the former a very interesting scenario to consider. This model has also been used to test various scenarios of super symmetric gravity \cite{Rosu} and quantum gravity, particularly in those that involve particular length scales, it was possible to impose bounds on those parameters of the theory \cite{Rosu,Castello,Pedram,M.Ruiz}.

In this work we showed that, for the case when there are no initial correlation between position and momentum, the effective Hamiltonian acquires quantum corrections related to the characteristic, ground-state energy and to the gravitational length $l_g$. The classical trajectory is bounded by the value of the time-dependent momentum $G^{0,2}$, because of the generalized uncertainty relation. It is worth mentioning that there are previous studies in the semiclassical regime for this quantum bouncer. For instance, in  \cite{Belloni} time-dependent solutions for different expectation values, for certain Gaussian states where found. It is interesting to note that they obtain a result very similar to ours for the dispersions in $x$ and $p$. However, since they do not consider the correlation between them, the product of $\Delta x \Delta p$ is a function of time, unlike our case, which is constant. The method of momentous effective quantum mechanics employed here results very suitable for the quantum bouncing ball for it provides a very detailed description of the evolution of the particle, which can be contrasted with experimental observations and measurements \cite{new,nat,seveso,owg,Benseghir}. The application of this method to the study of semiclassical states for more quantum systems, especially those with high experimental precision, will be carried out elsewhere.

%%%%%%%%%%%%%%%%%%%%%%%%%%%

\section*{Acknowledgments}

 Authors want to thank to the Organizer Committee of the {\it Tercer Encuentro de Modelado Matem\'atico en F\'isica y Geometr\'ia 2018} for the nice scientific environment that allowed the realization of this work. GCA want to thank Viviana Garc\'ia-Le\'on for useful discussions on the subject of this paper.

\end{document}